\documentclass{article}[16pt]
\usepackage[utf8]{inputenc}
\usepackage{graphicx}
\begin{document}

\begin{center}
{\Large \bf The Local Volume galaxy census and HST SNAP surveys}

\bigskip

{\large I.D.Karachentsev$^1$, M.I.Chazov$^1$, V.E.Karachentseva$^2$}

\end{center} 

   We examined F814W and F606W images of dwarf galaxies from the 
Hubble Space Telescope archive, which were obtained  under the HST 
SNAP programs 17159 to 17797. Among 58 observed dwarfs located outside the 
Local Group, we found only a few objects that were confidently resolved 
into stars.  We determined two new distances for the galaxies:
dw1252+2215  ($5.32\pm0.20$~Mpc) and dw1234+3952 ($4.34\pm0.16$~Mpc) 
via the Tip of the Red Giant Branch. They turned out to be new probable 
dwarf satellites of the nearby luminous spiral galaxies NGC\,4826 and 
NGC\,4736, respectively.  We also note that recent SNAP surveys vary in 
their productivity in measuring new galaxy distances by more than an 
order of magnitude.

  Key words:  {\em galaxies~--- dwarf galaxies, galaxies~--- distances}

 \bigskip 
One of the priority tasks of observational cosmology on small scales is the creation and replenishment of a representative sample 
of galaxies with measured radial velocities and accurate distances in the nearest fixed volume of the Universe. Analysis of deviations in the radial velocities 
of galaxies from the ideal (unperturbed) Hubble flow $V = H_0 \times D$, where $D$ is the distance to a galaxy and $H_0$ is the Hubble parameter, 
allows us to determine the mass of dark matter contained in local attractors. Radial velocities of galaxies measured by optical 
and radio telescopes have a characteristic accuracy of $\sim~(3-10)$~km~s$^{-1}$. 
Mass determinations of distances to nearby galaxies use a method based on a calibrated luminosity of the Tip of the Red Giant Branch (TRGB) [1, 2]. 
To apply this method, a stellar color-magnitude diagram based on images taken in F814W and F606W filters on the Hubble Space 
Telescope (HST) is used. The accuracy of the TRGB method is about 5\%, allowing one to record the position of a galaxy inside or outside a nearby group.

To date, the sample of galaxies in the Local Volume contains $\sim1500$ candidates with estimated distances within 12~Mpc [3]. 
For 550 of them, 
TRGB distances have already been measured and collected in the Extragalactic Distance Database [4]. More than a quarter of all TRGB
 distances in the Local Volume were obtained in the SNAP survey mode, when images of galaxies in both filters, F814W and F606W, were
 exposed during one HST orbital period in gaps between scheduled observations. Using such observational data, it is possible to measure TRGB
 distances up to $\sim12$~Mpc within one orbital period, that determines the size of the Local Volume.

During the last observing cycles on HST, two SNAP proposals were accepted for execution: SNAP 17158, ``The lowest luminosity
galaxy candidates ever discovered outside the Milky Way” (PI E.Bell, 55 orbits) and SNAP 17797 ``Galactic underdogs: assessing the true 
satellite galaxy population in the Local Void'' (PI E.Bell, 103 orbits). The purpose of these programs was to refine the luminosity function of 
galaxies at its faint end by measuring TRGB distances to satellites around nearby luminous galaxies (M\,31, M\,81, M\,64, M\,94, etc.), 
situated in the Local Volume.

The discovery of new satellites could resolve the well-known paradox that the expected number of satellites in the standard $\Lambda$CDM
 cosmological model with masses less than $10^{6-7}M_{\odot}$ is in the tens of times greater than their observed number [5, 6]. Successful searches 
for dwarf galaxies, especially of a low surface brightness, in modern deep sky surveys [7--10] have essentially reduced this
 contradiction.  Note that the weakening of this paradox also occurred due to the realization that a some satellites disappear in the process 
of evolution, being absorbed by massive host galaxies [11]. Nevertheless, a significant discrepancy between the expected and observed number 
of low-mass galaxies still remains.

We have carried out a visual inspection of images of all observed galaxies available in the HST archive for both the programs. Due to the circumstances of the
 SNAP surveys, images of only 20 galaxies were obtained within the framework SNAP\,17158 and 38 galaxies from SNAP\,17797. Typical
 exposures in F814W and F606W filters were about 750~sec. We excluded 18 putative satellites of M\,31 from consideration, since the
 images of these nearby objects with a scarce population require more sophisticated analysis. The color-magnitude diagrams for stars 
in four ultra-faint satellites of M\,81 have already been discussed by the authors of SNAP 17158 program [12]. These dwarf systems were previously
 discovered by ground-based observations with the 8-meter Subaru telescope [13--15]. New data from the SNAP survey [12] confirm the membership 
of these dwarfs in the M\,81 group, but do not lead to elaboration of their distances.

Based on our analysis of images of 20 galaxies from SNAP\,17158 and 33 galaxies from SNAP\,17797, we came to the following conclusions.
\begin{table} [h]
\caption{Properties of nearby dwarf galaxies}
\begin{tabular}{|ccccc|} \hline
 Parameter    &      Unit       &   dw1252+2215       &     dw1234+3952     &    dw0954+6821 \\
\hline
    TRGB      &        mag      &      24.60$\pm$0.08    &     24.11$\pm$0.08     &      24.00$\pm$0.08  \\  
    M(F814W)  &        mag      &          --4.08      &      --4.11          &            --4.10 \\
    $A_I$       &        mag      &           0.053      &       0.024          &              0.161 \\
    $(m-M)_0$    &        mag      &          28.17      &       28.19         &             27.94 \\
    Distance  &        Mpc      &        5.32$\pm$0.20   &       4.34$\pm$0.16    &        3.87$\pm$0.15 \\
    $g$         &        mag      &       19.86$\pm$0.13   &      18.46$\pm$0.10    &       (20.28)   \\
    $r$         &        mag      &       19.33$\pm$0.13   &      17.88$\pm$0.10    &          ---       \\ 
    $M_g$       &        mag      &       -8.9             &        -9.8            &       (-8.0)       \\          
  \hline
\end{tabular}
\end{table}              

 $\bullet$  For two galaxies, dw1252+2215 (J12\,52\,05.8 +22\,15\,54) and dw1234+3952 = SMDG\,1234+3952 (J12\,34\,22.4 +39\,52\,43), which are well resolved 
into stars, it is possible to determine their distances. 
The color-magnitude diagrams for them are shown in Figure~1. For photometry of the stellar population, the DOLPHOT package was used  [16]. 
The position of the Tip of the Red Giant Branch (TRGB) was estimated using the method developed by Makarov et al. [17].  The horizontal line in the 
figure shows the measured value of the TRGB. The results of the distance determination for 
both the galaxies are presented in Table~1.  In it, we have added a third well-resolved galaxy, dw0954+6821 (J09\,54\,07 +68\,21\,51), located 
in the M\,81 group, the CMD 
for which was already discussed in [12]. The rows of the table indicate: (1) the apparent magnitude of TRGB; (2) the absolute magnitude of TRGB, 
estimated from the calibrated relation
M(F814W) = --4.06 + 0.20 [$<$F606W--F814W$>$ -- 1.23],  justified in [2]; (3)  Galactic extinction in the $I$-band from [18]; (4,5)  the obtained 
distance modulus and linear distance of the galaxy with an indication of its standard error; (6,7) integrated $g$- and $r$-magnitudes of the 
galaxies according to [19]; in the case of dw0954+6821, the average integrated apparent magnitude is indicated according to the data of [15] 
and [12]; (8) the absolute magnitude of the galaxy.

 $\bullet$  The distance of the galaxy dw1252+2215 coincides within the measurement error with the distance of the luminous spiral galaxy NGC\,4826, $5.3\pm0.1$~Mpc, 
derived in [19]. Being at a projected separation  of $R_p = 116$~kpc, the dwarf galaxy is a new probable satellite of it. Note that Carlsten 
et al. [20]  give a similar distance of 5.42 Mpc for dw1252+2215 based on its surface brightness fluctuations.

 $\bullet$   The dwarf galaxy dw1234+3952 = SMDG1234+3952 is included in the Zaritsky catalog of ultra-diffuse galaxies [21]. Judging by its distance, 
this dwarf is a new satellite of the spiral galaxy M\,94 = NGC\,4736, which has a distance of $4.41\pm0.09$ Mpc [4]. The projection separation 
of both the galaxies is $R_p = 260$~kpc. It is curious that Zaritsky et al. [21] give an improbably large value of heliocentric radial velocity 
for the SMDG1234+3952, $V_h = 6658\pm16$~km~s$^{-1}$.

  $\bullet$  Among the other observed galaxies, the dwarf galaxy  dw1311+4051 appears to be the most resolved into stars. According to observations with 
the FAST radio telescope, it has a heliocentric radial velocity of $V_h = 604\pm1$~km~s$^{-1}$ [22] and a corresponding kinematic distance of 9.3~Mpc. 
The hydrogen mass of this dwarf,  $\log(M_{HI}/M_{\odot}) = 6.97$, is three times greater than the total mass of its stars.
The galaxy's CM-diagram shows a population of blue stars, which complicates the TRGB distance estimate. This dwarf is a probable satellite 
of the spiral galaxy NGC\,5055 with $D=9.04\pm0.09$~Mpc [4], being at a projection separation of $R_p = 229$~kpc from it.

\begin{figure}[hbt]
\includegraphics[height=10cm]{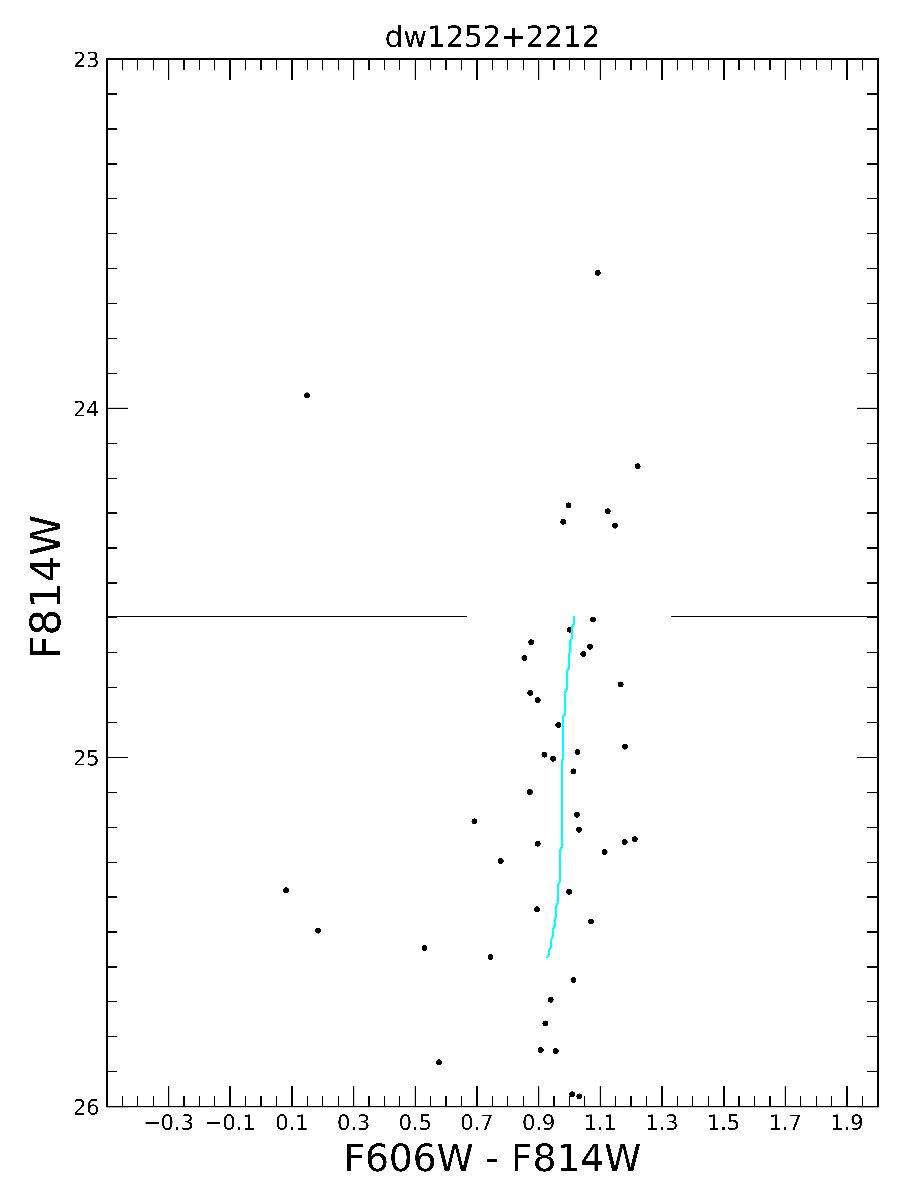}
\includegraphics[height=10cm]{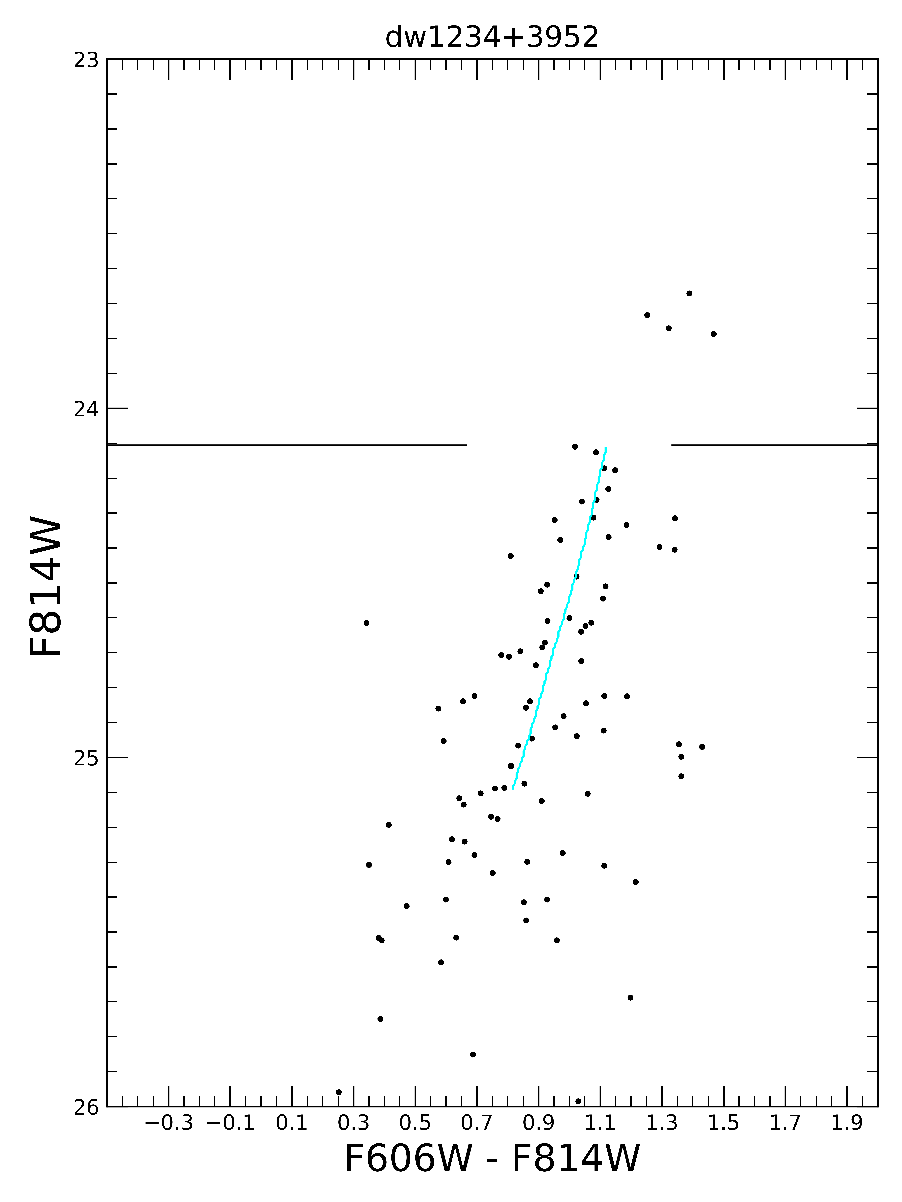}
\caption{Color-magnitude diagrams for stars in the galaxies dw1252+2215 and dw1234+3952.
The measured value of TRGB is shown by the horizontal line.}
  \end{figure}

  $\bullet$   Judging by the texture of the images resolved to extremely faint stars, 12 dwarf galaxies have distances $D < 12$~Mpc, i.e. are 
confirmed as the true members of the Local Volume sample. Of these, three objects: dw0852+3249, dw1046+1358 and dw1116+1356 are new satellites 
of galaxies NGC\,2683, NGC\,3377, and NGC\,3627, respectively.

 $\bullet$    Five dwarf galaxies represented in the UNGC catalog [4]: dw0315-4112, dw0321-4149, dw0846+3300, dw1116+5537 
and dw1122+5515 have a structure that indicates their belonging to the distant background, but not to satellites of massive 
galaxies in the Local Volume.

\begin{table} [h]
\caption{Five recent SNAP surveys aimed at measuring distances to nearby galaxies}
\begin{tabular}{|cccccc|} \hline
SNAP & PI. & Targets & Observed & NewDist & Efficiency \\ \hline
(1)& (2) & (3)& (4) & (5)& (6) \\
\hline
12546 & Tully & 132 & 45 & 42 &93\% \\
13442 & Tully & 128 & 45 & 43 &96\% \\
15922 & Tully & 153 & 80 & 53 &66\% \\
17158 & Bell & 55 & 20 & 2 &10\% \\
17797 & Bell & 103 & 33 & 1 &3\% \\ \hline

\end{tabular}
\end{table}

   It would be interesting to compare the performance of different SNAP programs in measuring new distances.
Table 2 contains the main findings from the five recent SNAP surveys that measured new distances of galaxies in the Local Volume. The first two
 columns indicate the name of the SNAP program and its Principal Investigator. The third and fourth columns represent the number of requested targets 
and the number of observed ones. The last two columns show the number of new galaxy distance estimates and, accordingly, the efficiency of the
 HST observations.

As one can see, the success in measuring new distances for galaxies in the Local Volume varies significantly among different SNAPs. The low efficiency of the last two programs
 is obviously due to a bad choice of objects whose distances could be determined with short exposures in a SNAP survey. Many targets in these
 programs were selected from Table 8 in the publication by Carlsten et al [20], where dubious candidates for satellites of nearby massive galaxies have 
been collected. The experience of the undertaken comparison may be useful in preparing subsequent SNAP proposals.

I.D.K. and M.I.C. acknowledge support from the Russian Science Foundation grant No.24--12--00277.

\

$^1$ Special Astrophysical Observatory of the Russian Academy of Sciences, Nizhny Arkhyz, Russia, e-mail: idkarach@gmail.com

$^2$ Main Astronomical Observatory, National Academy of Sciences of Ukraine, Kiev, 03143, Ukraine

\bigskip

{\large \bf References}

 1.  {\em M.G.Lee , W.L.Freedman , B.F.Madore}, Astrophys, J., {\bf 417}, 553, 1993.

 2. {\em L.Rizzi,  R.B.Tully, D.I.Makarov et al.,}  Astrophys. J., {\bf 661}, 815, 2007.
 
 3.  {\em I.D.Karachentsev, D.I.Makarov, E.I.Kaisina} Astron. J., {\bf145}, 101, 2013. (UNGC)

 4.  {\em G.S.Anand, L.Rizzi, R.B.Tully et al.,} Astron. J., {\bf 162}, 80, 2021.
 
 5.  {\em A.Klypin, A.V.Kravtsov, O.Valenzuela, F.Prada}, Astrophys. J.,  {\bf 522}, 82, 1999.
 
 6. {\em B.Moore, S.Ghigna, F.Governato et al.,}  Astrophys. J., {\bf 524}, L19, 1999.
 
 7. {\em K.N.Abazajian, J.K.Adelman-McCarthy, M.A.Agueros  et al.,}  Astrophys. J. Suppl., {\bf 182}, 543, 2009.
 
 8.  {\em A.Dey, D.J.Schlegel, D.Lang et al.,} Astron. J., {\bf 157}, 168, 2019.
 
 9.  {\em C-P.Zhang, M.Zhu, P.Jiang  et al.,} Science China Physics, Mechanics and Astronomy, {\bf 67}, 219511, 2024.
 
10. Euclid Collaboration, Astronomy and Astrophys.,  {\bf 691A}, 175, 2024.

11.  {\em Y.Kim,  A.H.G.Peter, J.R.Hargis et al.,}  Physical Review Letters, {\bf 121u}, 1302, 2018.

12.  {\em K.Gozman, E.Bell, I.S.Jang et al.,} Astrophys.J., {\bf 977}, 179, 2024.

13.  {\em A. Smercina,  E.Bell, C.T.Slater  et al.,}  Astrophys.J., {\bf843L}, 6, 2017.

14. {\em S.Okamoto, N.Arimoto, A.M.N.Ferguson  et al.,} Astrophys. J.,  {\bf884}, 128, 2019.

15. {\em E.Bell, A.Smercina, P.A.Price  et al.,} Astrophys. J. Lett., {\bf937L}, 3, 2022.

16. {\em A.E.Dolphin}, Mon.~Not.~Roy.~Astr.~Soc., {\bf332}, 91, 2002.

17. {\em D. Makarov, L. Makarova, L. Rizzi, et al,} Astron. J., {\bf132}, 2729, 2006

18. {\em E.F.Schlafly, D.P.Finkbeiner},  Astrophys.J., {\bf737}, 103, 2011.

19. {\em J. Mould, S. Sakai}, Astrophys.J. Lett., {\bf686}, L75, 2008

20. {\em S.G.Carlsten, J.E.Greene, R.L.Beaton et al.,} Astrophys. J., {\bf933}, 47, 2022.
 
21. {\em D.Zaritsky,  R.Donnerstein, A.Dey, et al.,} Astrophys.J. Suppl, {\bf267}, 27, 2023.

22.  {\em I.D. Karachentsev, V.E. Karachentseva, S.S. Kaisin, C.-P. Zhang} Astron. \& Astrophys., {\bf684L}, 24, 2024.

\clearpage

\end{document}